\documentstyle[psfig,natbib]{elsart}
\def\astrobj#1{#1}
\def\url#1{{\ttfamily\def\/{/\discretionary{}{}{}}#1}}

\begin{document}

\begin{frontmatter}
\title{A new measurement of zinc metallicity in a DLA at z $\sim$
3.35$^{1}$}

\author[IoA]{C\'eline P\'eroux}
\author[IAP,ObsParis]{Patrick Petitjean}
\author[IAP]{Bastien Aracil}
\author[Pune]{R. Srianand}

\address[IoA]{Institute of Astronomy, Madingley Road, Cambridge CB3
0HA, UK.}

\address[IAP]{Institut d'Astrophysique de Paris, 98bis Boulevard
Arago, Paris, France.}

\address[ObsParis]{LERMA, Observatoire de Paris, 61, avenue de
l'Observatoire, F-75014, Paris, France.}

\address[Pune]{IUCAA, Post Bag 4, Ganeshkhind, Pune 411 007, India.}

% use the thanksref command within \title, \author or \address for footnotes:
% \title{\thanksref{*}}
  \thanks[]{Based on observations collected during programme ESO
65.P-0038 and ESO 69.B-0108 at the European Southern Observatory with
UVES on the 8.2m KUEYEN telescope operated at the Paranal Observatory,
Chile}
% \author{\thanksref{label2}}
% \thanks[label2]{}
% \address{\thanksref{label3}}
% \thanks[label3]{}
% including your email address
% \address{\thanksref{email}}
 \thanks[email]{E-mail: celine@ast.cam.ac.uk}

\begin{abstract}
We present chemical abundance measurements in the $z_{abs}=3.35045$
Damped Lyman-$\alpha$ (DLA) system observed in the UVES spectrum of
the BAL quasar \astrobj{BR 1117$-$1329}. We measure a neutral hydrogen column
density $N (H I) =6.9 \pm 1.7 \times 10^{20}$ atoms cm$^{-2}$ and
derive mean abundances relative to solar: $[Si/H] = -1.26\pm0.13$,
$[Fe/H]=-1.51\pm0.13$, $[Ni/H]=-1.57\pm0.13$, $[Cr/H]=-1.36\pm0.13$,
$[Zn/H]=-1.18\pm0.13$, $[Al/H]>-1.25$, $[O/H]>-1.25$ and
$[N/H]<-2.24$. This is the third measurement of Zn, an element mildly
depleted onto dust grain, at $z_{abs}>3$. The iron to zinc and
chromium to zinc ratios, $[Fe/Zn]=-0.33\pm0.05$ and
$[Cr/Zn]=-0.18\pm0.05$ demonstrate that the absorber has a low dust
content. The nitrogen ratio $[N/Si]<-0.98$ suggests that the
``secondary'' N production process is taking place in this
DLA. Finally, this absorber does not seem to present a convincing
$\alpha$-enhancement as shown by the $\alpha$ over Fe-peak element
ratios: $[Si/Fe]=0.25\pm0.06$, $[Si/Cr]=0.10\pm0.06$ and
$[Si/Zn]=-0.08\pm0.06$.
\end{abstract}

\end{frontmatter}

\section{Introduction}

The chemical abundances of the highest H I column density quasar
absorbers ($N(HI)> 2 \times 10^{20}$ atoms cm$^{-2}$), the Damped
Lyman-$\alpha$ systems (DLAs), are commonly expressed using column
density weighted measurements. They are used as observational tracers
of the cosmological evolution of metallicities, up to very high
redshifts (see for example Pettini et al. 1994, 1997; Lu et al.  1996;
Prochaska \& Wolfe 1999, 2000, 2001; Dessauges-Zavadsky \etal\ 2001;
Ledoux, Bergeron \& Petitjean 2002 and references therein).  However,
recent studies give somehow surprising results: Prochaska \& Wolfe
(2002) claim no evolution of the mean weighted iron metallicity over
the redshift range $1.7-3.5$, in contradiction with predictions from
essentially all chemical evolution models. A possible explanation for
this result is that iron is not the optimal element for tracing metals
on cosmological scales, because it is easily depleted onto dust
grains.  Dust depletion greatly complicates the interpretation of the
metal content of DLAs. Nevertheless at very high-redshift ($z>3.8$),
there are evidences that [Fe/H] is beginning to fall, in such a way
that at $z=5$ the metallicity is substantially lower than at $z<4$
(Prochaska, Gawiser \& Wolfe 2001, Songaila \& Cowie 2002).

\begin{figure}
\psfig{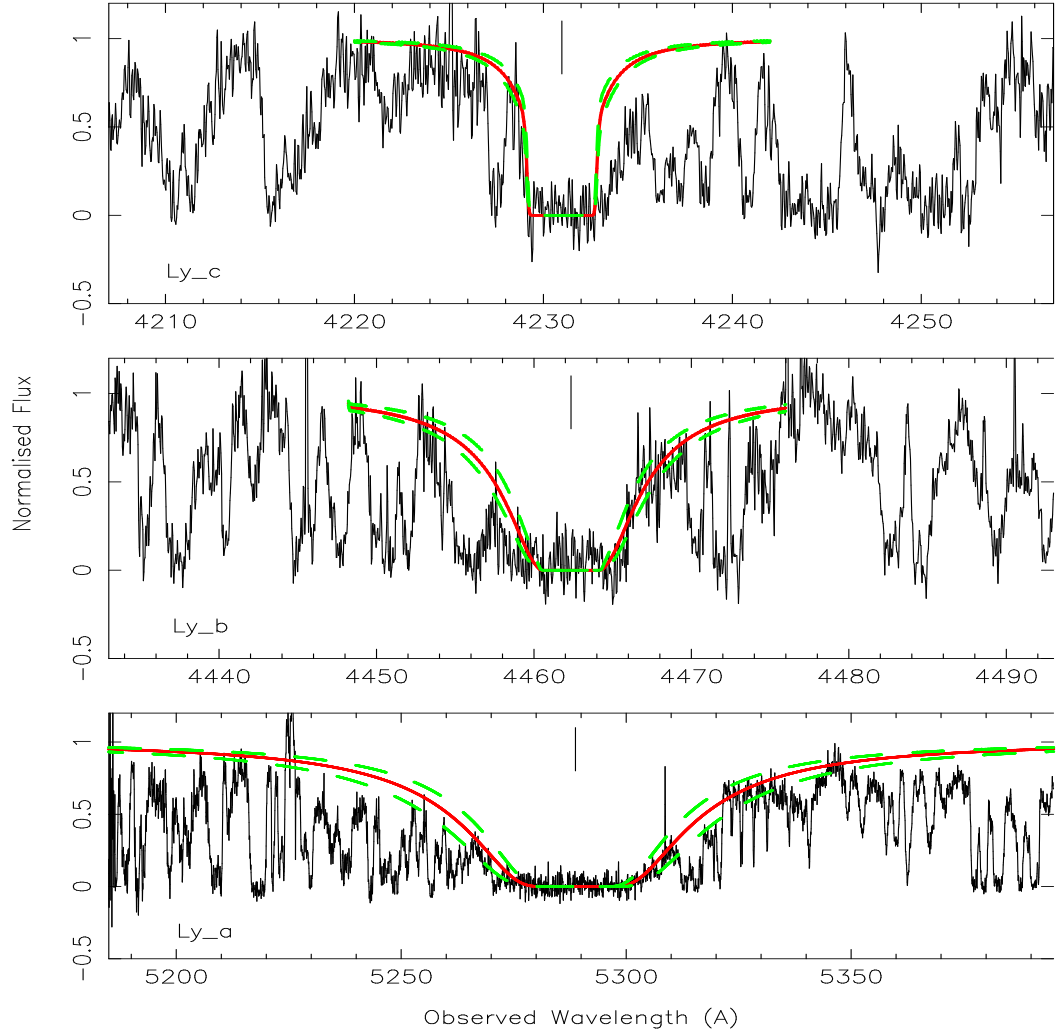}
\caption{ Voigt profile fits to the Lyman series lines (Ly-$\alpha$,
Ly-$\beta$ and Ly-$\gamma$) of the $z=3.35045$ DLA. The vertical bar
in each panel indicates the wavelength centroid of the component used
for the best fit, which is shown as a solid line together with 1
$\sigma$ errors (dashed lines). The resulting total HI column density
is $N$(HI)$=6.92 \pm 1.7 \times 10^{20}$ atoms cm$^{-2}$.}
\label{fig:DLA}
\end{figure}

Pettini \etal\ (1994) estimated the dust content of DLAs with the help
of a comprehensive survey of Zn measurements, an element which is
known to be only slightly depleted onto dust grains, and thus provide
an unbiased tracer of metallicities. However such measurements are
challenging because of both the paucity and weakness of Zn features in
quasar absorbers. At the moment, only two measurements of [Zn/H] have
been made in DLAs at z$>$3 (Molaro \etal\ 2000, Levshakov \etal\
2001). Here, we present a third detection of Zn in a DLA at
z$\sim$3.35, thus providing clues on the dust-free abundance
determination in high-redshift Damped Lyman-$\alpha$ systems.  We
first detail the observational set-ups and data reduction processes of
the quasar spectrum. We then present the analysis and chemical
abundance determination for a number of elements in the DLA studied,
providing a short discussion on the consequences of these additional
measurements at high-redshift.

\section{Observation and Data Reduction}

\begin{figure}
\psfig{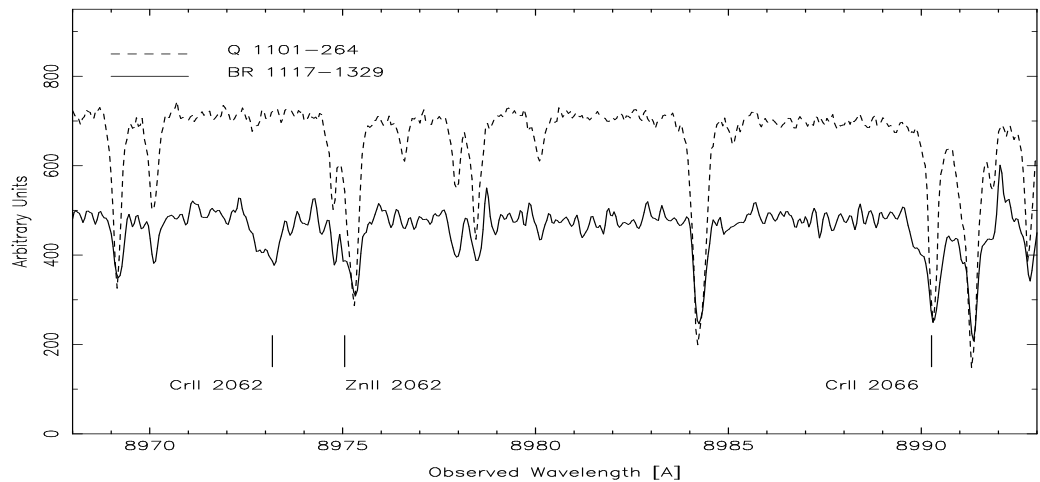}
\caption{Comparison between the spectrum of BR~1117$-$1329 (solid) and
that of another high-redshift quasar taken as a reference
(dashed). The superposition of the two spectra allows to check for
possible atmospheric contamination in the observed wavelength range
8968\AA--8993\AA. The Zn II $\lambda$ 2062 and Cr II $\lambda$ 2066
are contaminated by sky features while the Cr II $\lambda$ 2062 line
appears clear from any contamination. }
\label{fig:sky}
\end{figure}

\astrobj{BR 1117$-$1329} is a high-redshift ($z$ $\sim$ 3.96) quasar exhibiting
broad absorption lines (BAL) discovered by Storrie-Lombardi et
al. (1996). The data presented in this study were obtained using the
Ultraviolet-Visual Echelle Spectrograph (UVES) on the 8.2m VLT in
April 2000 and April 2002. UVES is particularly well suited for the
study of Zn at high-redshift thanks to its good sensitivity at the
longer wavelengths. Three partially overlapping setups (480+710+725)
were used to provide a complete coverage of the quasar spectrum
ranging from $\sim$ 4195 \AA\ to 9200 \AA.

The individual spectra were reduced using the UVES data reduction
pipeline of the ESO MIDAS package (see Ballester \etal\ 2000 for a
detailed description). The steps of the data processing include
wavelength calibration, order extraction and flat-fielding. Once
extracted, the individual frames were corrected to a vacuum
heliocentric scale and combined together resulting in a spectrum with
signal-to-noise ratio of about 40 per pixel at $\sim$ 7000\AA. The
quasar continuum was then fitted to the quasar spectrum using a spline
function.

\begin{figure}
\psfig{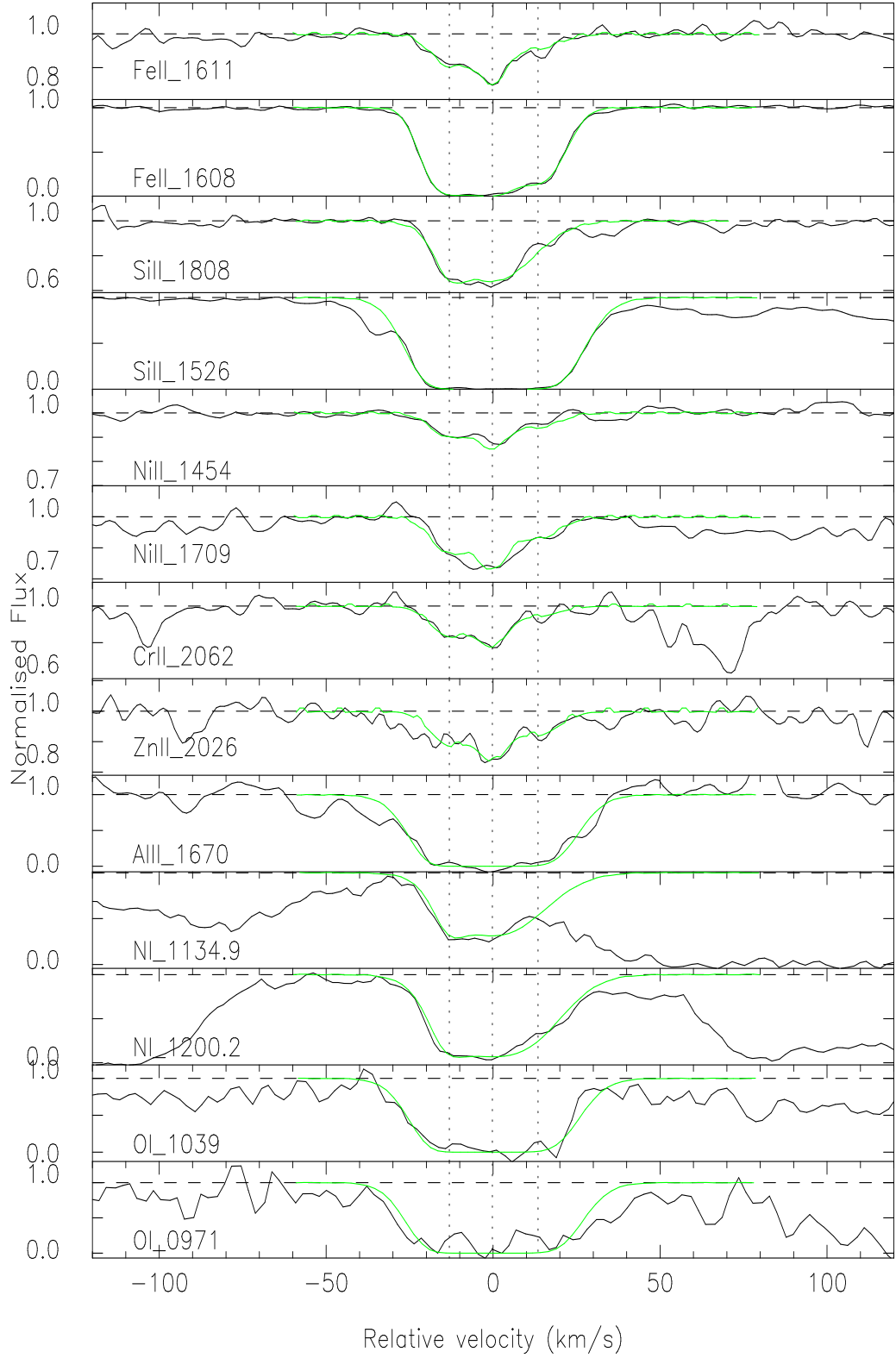}
\caption{Velocity profiles (km s$^{-1}$) of the metals in the DLA
absorption system with the fit overplotted (light colored solid
line). The spectra are normalized to unity. The zero velocity is fixed
at $z_{abs}=3.35045$ and the dashed lines indicate the positions of
the 3 components.}
\label{fig:low_ioni1}
\end{figure}

A careful inspection of the spectrum bluewards of the Lyman-$\alpha$ emission
reveals a DLA absorption in a region free of BAL features. The redshift
($z_{abs}=3.35045$) is well defined by the presence of 
higher Lyman series lines and  heavy element absorption lines including
some from Zn~{\sc ii}.

\section{Analysis}
        
\subsection{Column Densities Measurements}

The H~{\sc i} column density of this system is measured by fitting a
theoretical Voigt profile to the Lyman series. The quasar continuum
was fitted with a spline function in MIDAS. As always in DLAs at
high-redshift, the positioning of the continuum over the strong
absorption lines is the main source of uncertainties in the H~{\sc i}
column density measurement. The use of several lines to perform the
fit gives confidence in the result however.  The fitting of the
profile is performed with VPFIT~\footnote{VPFIT was written by
R.F.Carswell, J.K. Webb, M.J. Irwin, A.J. Cooke. More information
about VPFIT are available at following URL:
http://www.ast.cam.ac.uk/$\sim$rfc/vpfit.html} (Webb \etal\ 1987) and
the resulting fit is presented in Figure 1 for the first three lines
of the Lyman series. The total H~{\sc i} column density is $N$(H~{\sc
i}) $=6.92 \pm 1.7 \times 10^{20}$ atoms cm$^{-2}$.

We derive the column density of the ions associated with the DLA by
fitting theoretical Voigt profiles to the observed features minimizing
$\chi^2$. For this purpose we make use of the FITLYMAN package in the
MIDAS data reduction software. Only the lines free of telluric
contamination and blending are used to determine the column densities
of the elements. In particular, we took great care to check possible
atmospheric contamination in the Cr~{\sc ii} $\lambda$ 2062 region
(see Figure~\ref{fig:sky}).

\begin{table}
\begin{center}
\caption{Summary table of the oscillator strength values used in this
study.}
\label{tab:fosc}
\begin{tabular}{l c l ll }
\hline\hline
Ion & $f_{osc}$ &References \\
\hline
Si~{\sc ii} $\lambda$ 1526 	&0.11000 	&Spitzer \& Fitzpatrick (1993)\\Si~{\sc ii} $\lambda$ 1808      &0.00218	&Bergeson \& Lawler (1993b)\\
Fe~{\sc ii} $\lambda$ 1608	&0.05800	&Bergeson \etal\ (1996b)\\    
Fe~{\sc ii} $\lambda$ 1611	&0.00130 	&Welty \etal\ (1999)\\     
Ni~{\sc ii} $\lambda$ 1454	&0.03230	&Fedchak \etal\ (2000)\\     
Ni~{\sc ii} $\lambda$ 1709	&0.03240	&Fedchak \etal\ (2000)\\     
Cr~{\sc ii} $\lambda$ 2062	&0.07796	&Bergeson \& Lawler (1993a)\\  
Zn~{\sc ii} $\lambda$ 2026	&0.48860	&Bergeson \& Lawler (1993a) \\ 
Al~{\sc ii} $\lambda$ 1670	&1.83300	&Morton (1991) \\
O I $\lambda$ 971		&0.01214	&Morton (1991) \\
O I $\lambda$ 1039		&0.00920	&Morton (1991)\\
%N I $\lambda$ 1134.9		&0.04023	&Morton (1991)\\
N I $\lambda$ 1200.2		&0.08849	&Morton (1991)\\
C~{\sc iv} $\lambda$ 1548	&0.19080	&Morton (1991)\\
C~{\sc iv} $\lambda$ 1550	&0.09522	&Morton (1991)\\
Al~{\sc iii} $\lambda$ 1854	&0.56020	&Morton (1991)\\
Al~{\sc iii} $\lambda$ 1862	&0.27890	&Morton (1991)\\
\hline
\hline
\end{tabular}
\end{center}
\end{table}
\vspace{1cm}

\subsection{Results}

The column densities obtained from the Voigt profile fits are used to
derive abundances of Fe, Si, Ni, Zn, Cr and to place a lower limit on
the abundance of Al and O and an upper limit on the abundance of
N. All the low ionization species were fitted together with 3
components (see Table~\ref{tab:low}).  We notice that all the lines
can not be fitted with a single Doppler parameter. However a  

\newpage
\begin{table}
\begin{center}
\caption{Results of Voigt profile fits to low ionization species in
the z = 3.35045 DLA system.}
\label{tab:low}
\begin{tabular}{l c l l c }
\hline\hline
Component & $z_{\rm abs}$ & Ion & $b^{a}$  & $\log$ N(X)$^{b}$  \\
     &               &     & [km/s]   & [cm$^{-2}$]      \\     
\hline
A    & 3.350265 & FeII 1608--1611   & 6.5$\pm$0.5   & 14.39$\pm$0.01  \\     
     &          & SiII 1526--1808   & 4.0$\pm$0.4   & 14.34$\pm$0.05  \\      
     &          & NiII 1454--1709   & 6.5$\pm$0.5   & 13.07$\pm$0.02  \\
     &          & CrII 2062   & 6.5$\pm$0.5   & 12.76$\pm$0.02  \\      
     &          & ZnII 2026   & 6.5$\pm$0.5   & 11.79$\pm$0.03  \\      
     &          & AlII 1670   & ...   & $>$13.85$^{c}$         \\
     &          & OI 971--1039& ...   & $>$15.85$^{c}$         \\
     &          & NI 1200.2   & ...   & $<$13.75$^{c}$         \\
B    & 3.350452 & FeII 1608--1611   & 4.9$\pm$0.7   & 14.50$\pm$0.03  \\     
     &          & SiII 1526--1808   &14.8$\pm$0.1   & 15.04$\pm$0.05  \\      
     &          & NiII 1454--1709   & 4.9$\pm$0.7   & 13.14$\pm$0.02  \\
     &          & CrII 2062   & 4.9$\pm$0.7   & 12.80$\pm$0.02  \\      
     &          & ZnII 2026   & 4.9$\pm$0.7   & 11.86$\pm$0.02  \\      
     &          & AlII 1670   &...   & $>$13.65$^{c}$  	 \\
     &          & OI 971--1039&...   & $>$16.15$^{c}$         \\
     &          & NI 1200.2   &...   & $<$14.39$^{c}$         \\
C    & 3.350650 & FeII 1608--1611   & 7.7$\pm$0.5   & 14.09$\pm$0.01  \\     
     &          & SiII 1526--1808   & 15.0$\pm$0.1   & 13.51$\pm$0.05  \\      
     &          & NiII 1454--1709   & 7.7$\pm$0.5   & 12.88$\pm$0.04  \\
     &          & CrII 2062   & 7.7$\pm$0.5   & 12.30$\pm$0.05  \\      
     &          & ZnII 2026   & 7.7$\pm$0.5   & 11.67$\pm$0.04  \\      
     &          & AlII 1670   &...   & $>$11.80$^{c}$  	 \\
     &          & OI 971--1039&...   & $>$14.50$^{c}$         \\
     &          & NI 1200.2   &...  & $<$13.55$^{c}$         \\
\hline
\hline
\end{tabular}
\end{center}
\end{table}

%\vspace{1.0cm} 

$^a$ The good fit is achieved by assuming two different Doppler values
respectively for Fe, Ni, Zn, Cr and Si, Al, O, N. The latter set of
lines is broader than the other one.  This difference is possibly due
to the presence of extra components in these lines.  \\
$^b$ Limits are 3-$\sigma$.  \vspace{.5cm}$^{c}$ Using same b-parameter as in Si~{\sc ii}.\\

\vspace{.5cm}      	       

good fit is achieved by assuming two different values respectively for
Fe, Ni, Zn, Cr and Si, Al, O, N. The latter set of lines are slightly
broader than the other one.  This difference is possibly due to the
presence of extra components. It must be noted that this does not
affect the column density determination as most of the species have at
least one transition which is optically thin. An exception may be
Si~{\sc ii} for which the weakest observed transition Si~{\sc ii}
$\lambda$1808 is partially saturated (although not going to
zero). Errors indicated in the tables are statistical errors in the
fit. However, there are also systematic errors not accounted for here
but which are smaller than the statistical errors.

The metal content with respect to solar values is expressed in the
usual manner: $[X/H] = \log [N(X)/N(H)]_{DLA}- \log
[N(X)/N(H)]_{\odot}$, assuming that $N(X) = N (X II)$ and $N (H) = N
(H I)$, where the solar values are taken from Grevesse \& Savage
(1998) and Holweger (2001; see Table~\ref{tab:summary}).  Silicon
abundance is derived from the slightly saturated (but not going to
zero) Si~{\sc ii} $\lambda$ 1808 transition and the saturated Si~{\sc
ii} $\lambda$ 1526 transition.  This results in a [Si/H] value of
$-1.26\pm0.13$. The error in the Si~{\sc ii} column density also
includes the uncertainty in b's determination. Similarly the iron
abundance is obtained from the optically thin Fe~{\sc ii} $\lambda$
1611 and saturated Fe~{\sc ii} $\lambda$ 1608 transitions. We derive
$[Fe/H]=-1.51\pm0.13$. Two Ni~{\sc ii} lines (Ni~{\sc ii}
$\lambda\lambda$ 1454, 1709) are available for abundance
determination. We measure the column density:
$[Ni/H]=-1.57\pm0.13$. The Cr~{\sc ii} lines are detected (Cr~{\sc ii}
$\lambda$$\lambda$$\lambda$ 2056,2062,2066), but only one line is
suitable for abundance determination (see Figure~\ref{fig:sky}). The
abundance derived from the Cr~{\sc ii}$\lambda$ 2062 is
$[Cr/H]=-1.36\pm0.13$. After checking for possible atmospheric
contaminations at these wavelengths, we find the Zn~{\sc ii} $\lambda$
2026 line allows for the determination of the Zn content of this
DLA. The new measurement of Zn at $z_{abs}>3$, $[Zn/H]=-1.18\pm0.13$,
is higher than other measurements at these redshifts. The Al~{\sc
ii}$\lambda$ 1670 line is heavily saturated and hence only a lower
limit can be derived from this absorption feature:
$[Al/H]>-1.25$. Similarly, several O~{\sc i} lines are detected but
all are saturated. Using O I $\lambda$ 971 and OI $\lambda$ 1039, we
derive: $[O/H]>-1.25$. Finally, the two red most triplets of N~{\sc i}
are observed but are severely blended. Only one of the lines at our
disposal is suitable to determine an upper limit to the N I abundance
in this system: N I $\lambda$ 1200.2. We derive: $[N/H]<-2.24$.

\begin{figure}
\psfig{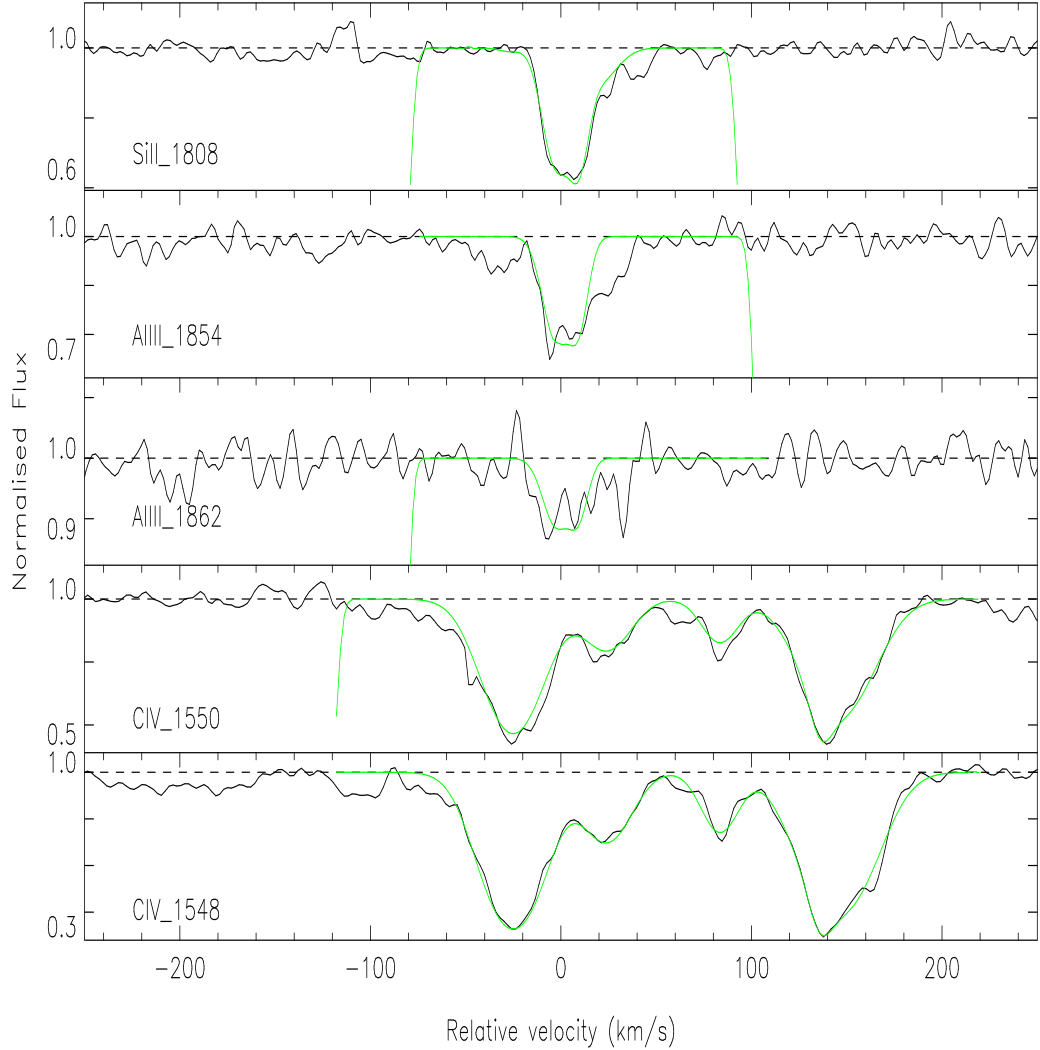}
\caption{Velocity profiles (km s$^{-1}$) of heavy element lines in the
DLA absorption system with the fit overplotted (light colored solid
curve). The spectra are normalized to unity. The zero velocity is
fixed at the redshift of the absorption feature: $z_{abs}=3.35045$.}
\label{fig:high_ioni}
\end{figure}

\begin{table}
\begin{center}
\caption{Results of Voigt profile fits to high-ionization species in
the z = 3.35045 DLA system.}
\label{tab:high}
\begin{tabular}{l c l lll }
\hline\hline
Component & $z_{\rm abs}$ &Vel& Ion & $b_{thermal}$  & $\log$ N(X)  \\
     &               &&    & [km/s]    & [cm$^{-2}$]      \\     
\hline
A    &3.350745  &$-$25     &CIV    &20.9$\pm$0.4  &13.73$\pm$0.01\\     
B    &3.351459  &$+$24     &CIV    &16.6$\pm$0.9  &13.14$\pm$0.02\\     
C    &3.352319  &$+$83     &CIV    &12.4$\pm$1.0  &12.95$\pm$0.02\\     
D    &3.353086  &$+$136    &CIV    & 6.0$\pm$1.2  &12.78$\pm$0.07\\     
E    &3.353214  &$+$145    &CIV    &24.5$\pm$0.4  &13.73$\pm$0.01\\     
a    &3.351072  &$-$6      &AlIII  & 8.5$\pm$2.4  &12.17$\pm$0.02\\      
b    &3.351246  &$+$7      &AlIII  & 5.2$\pm$1.4  &11.92$\pm$0.03\\      
\hline
\hline
\end{tabular}
\end{center}
\end{table}

\begin{figure}
\psfig{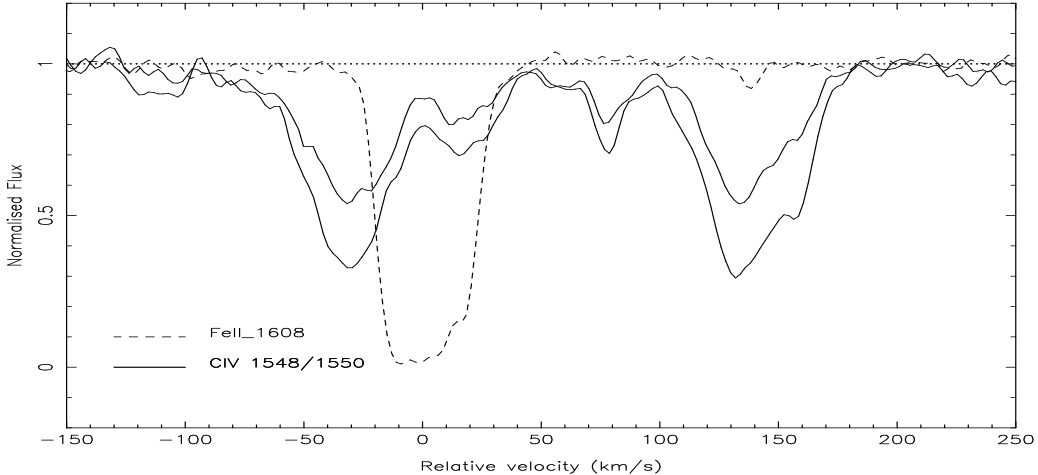}
\caption{Fe II $\lambda$ 1608 and the C IV doublet ($\lambda$ 1548 and
$\lambda$ 1550\AA) overplotted on the same velocity scale. This figure
illustrates the shift between the center of the low and the high
ionization ionic features in the DLA. The zero velocity is fixed at
the redshift of the absorption feature: $z_{abs}=3.35045$.}
\label{fig:overplot_ions}
\end{figure}

We also fit the high-ionization species in the DLA. The resulting
column density estimates are summarized in Table~\ref{tab:high}
together with the redshift and Doppler parameters associated with each
individual component. 
The C~{\sc iv} doublet (C~{\sc iv}$\lambda$$\lambda$1548,1550) is
spread over $\sim$ 300 km s$^{-1}$. It is well fitted by 5 components
(Figure~\ref{fig:high_ioni}). We note that there is a velocity shift
between the centroid of the low-ionization line profiles (Si~{\sc ii}
$\lambda$ 1808, Fe~{\sc ii} $\lambda$ 1608, etc) and the one from the 
C~{\sc iv} doublet. This is illustrated in Figure~\ref{fig:overplot_ions} 
where the Fe~{\sc ii}$\lambda$1608 line profile is plotted together with 
the two members of the C~{\sc iv} doublet on the same velocity scale.
    
Finally, the Al~{\sc iii} doublet (Al~{\sc
iii}$\lambda$$\lambda$1854,1862) is fitted with two components which
have their positions and $b$-values fixed to those of Si~{\sc
ii}. Indeed, the velocity structure of Al~{\sc iii} is known to follow
closely the one from singly ionized species in DLAs (Lu \etal\ 1996,
Prochaska \& Wolfe 1999, Prochaska \& Wolfe 2000). This observational
fact is not straight forward to explain: the ionization potential of
Al~{\sc ii} being greater than the one of hydrogen, Al~{\sc iii} is
not expected to be present in the regions dominated by neutral
hydrogen (i.e. DLA) although the atomic physics of aluminum species is
not well known (see Petitjean et al. 1994). Several studies have tried
to explain the observations (Howk \& Sembach 1999, Izotov, Schaerer \&
Charbonnel 2001, Vladilo \etal\ 2001).

\section{Discussion and Conclusion}

In the DLA studied here, we observe a velocity shift between the low
and the high ionization ionic features. The fact that the
high-ionization profiles show much more disturbed velocity structure
than the low-ionization profiles has already been noticed in the past
(Lu \etal\ 1996; Ledoux \etal\ 1998). Haehnelt, Steinmetz \& Rauch
(1998) have suggested that such feature could be the signature of
merging protogalactic clumps. Indeed they use numerical simulations to
model the Si~{\sc ii}$\lambda$1808 and C~{\sc iv}$\lambda$1548 line
profiles and note that the absorption features vary independently in
the high ionization and low ionization species since the C~{\sc iv}
absorption arises mainly from the warmer gas outside the
self-shielding region of DLAs. Nevertheless, other numerical
simulations fail to reproduce the DLA kinematics in a self-consistent
manner (e.g. Prochaska \& Wolfe 2001) and actually require that a
significant fraction of DLAs have $v_{circ} \sim$ 150 km s$^{-1}$
(Maller \etal\ 2001) even within the Cold Dark Matter cosmology. In
other words, if one wants to interpret the global kinematics of the
whole DLA population with a simple model, the kinematics of the
strongest systems might rule out a global dwarf galaxy scenario if one
adopts $v_{circ} <$ 100 km s$^{-1}$ as the dwarf
criterion. Nevertheless, note that in the case of BR~1117$-$1329, the
neutral component has a velocity width of about 50~km~s$^{-1}$ and
that the strongest C~{\sc iv} components are seen on both side (at
$-$50 and +130~km~s$^{-1}$) of the low-ionization profile. This is
suggestive of a galactic wind from a dwarf galaxy.

The detection of N~{\sc i} in this DLA system allows us to probe the
chemical history of the system. Indeed, the [N/O] ratio is useful to
disentangle the contribution from the ``primary'' and ``secondary''
production processes for N enrichment.  The limits obtained here
suggests that [N/O]$\le-0.9$ which is slightly lower than what one
expects (i.e., $\simeq-0.6$) if the primary process dominates
(Vila-Costas \& Edmunds, 1993). 
%If one assumes [O/H] is very much
%similar to [Si/H] and silicon is not strongly depleted onto dust
%grains then it is expected that [N/O] is $\sim-1.2$ purely from the
%secondary production. 
Like in most of the DLAs the [O/N] limits for the measured
$\alpha-$element metallicity is consistent with the delayed production
of N compared to O and inconsistent with the primary production of N
from the massive stars in the low metallicity gas (Lipman, Pettini \&
Hunstead, 1995; Lu \etal\, 1998; Centuri\'on \etal\ 1998; Pettini
\etal\ 2002 {\it in press}; Prochaska \etal\, {\it in press}).

The abundances of the DLA have been studied in detail. In this DLA,
we find a zinc abundance slightly higher than the iron one:
$[Fe/Zn]=-0.33\pm0.05$ which suggests that the amount of dust present
in this absorber is rather low. This is further supported by the
chromium over zinc ratio $[Cr/Zn]=-0.18\pm0.05$, another tracer of
dust. Thus in this DLA the abundances of refractory elements (such as
Fe and Cr) are in line with the one from the non-refractory element
Zn. In addition, Al~{\sc iii}/Al~{\sc ii} is less than 5\%. This means
that the effect of ionization correction in the measurement of
metallicity in this system is almost negligible (Vladilo et al. 2001)
and the absolute abundances derived here do not suffer severe biases
due to dust depletion.

We derive various $\alpha$-chain element to Fe-peak element ratios:
$[Si/Fe]=0.25\pm0.06$, $[Si/Cr]=0.10\pm0.06$ and
$[Si/Zn]=-0.08\pm0.06$.  Thus the abundance pattern in this system is
consistent with no $\alpha-$enhancement with very little depletion in
Fe and Cr with respect to Zn. The results indicate that the
Type~II~SNe enrichment process does not dominate in this system,
possibly because the Type~I~SNe enrichment process is already
important. The fact that we observe the DLA when Type~I~SNe processes
start to dominate over Type~II~SNe can not easily be interpreted since
the turn-over depends on both the star formation rate and the initial
mass function (Matteucci \& Recchi 2001).

\begin{table}
\begin{center}
\caption{Summary table of the column densities of the low and high
ionization species in the $z_{abs}=3.35045$ DLA.}
\label{tab:summary}
\begin{tabular}{lclc}
\hline
\hline
Ion	&$\log$ [N(X)/N(H)]$_\odot$ (Ref)&$\log$ N(X)&[X/H]\\
\hline
H I 	 &...		&20.84$\pm$0.12 &...\\
Fe II    &$-$4.50 (1)	&14.83$\pm$0.03	&$-$1.51$\pm$0.13\\
Si II    &$-$4.45 (1)	&15.13$\pm$0.05	&$-$1.26$\pm$0.13\\
Ni II    &$-$5.75 (1)	&13.52$\pm$0.03	&$-$1.57$\pm$0.13\\
Cr II    &$-$6.33 (1)	&13.15$\pm$0.03	&$-$1.36$\pm$0.13\\
Zn II    &$-$7.40 (1)	&12.26$\pm$0.03	&$-$1.18$\pm$0.13\\
Al II    &$-$5.53 (1)	&$>$14.06      	&$>-$1.25\\
OI	 &$-$3.26 (2)   &$>$16.33	&$>-$1.25\\
NI	 &$-$4.07 (2)	&$<$14.53	&$<-$2.24\\
C IV     &...    	&14.13$\pm$0.01	&...\\	 
Al III   &...		&12.36$\pm$0.03	&...\\
\hline
\hline
\end{tabular}
\vspace{.5cm}      	       

References:\\

[1] Grevesse \& Savage 1998.\\

[2] Holweger 2001.\\

\end{center}
\end{table}

We derive a Zn abundance, $[Zn/H]=-1.18\pm0.13$ at $z_{abs}=3.35$,
which is higher than the very few measurements available at these
redshifts (corrected to the same solar value as use in the present
study): Molaro \etal\ (2000) derives $[Zn/H]=-2.00\pm0.1$ in a DLA at
$z_{abs}=3.39$ towards QSO 0000$-$2620 and Levshakov \etal\ (2002)
find $[Zn/H]=-1.43\pm0.08$ in an absorber at $z_{abs}=3.02$ towards
QSO 0347$-$3819 although the metallicity in this absorber is uncertain
and could be larger (see Ledoux, Srianand \& Petitjean 2002). The
resulting column density weighted mean at $z>3$ is
$[<Zn/H_{DLA}>]=-1.63$, i.e. dominated by the highest HI column
density, which here corresponds to the lowest Zn measurement from
Molaro \etal\ (2000). For comparison, Pettini \etal\ (1997) find
$[<Zn/H_{DLA}>]=-1.18$ from $z=0.69$ to $z=3.39$ (where the measures
at $z>3$ are upper limits). Although statistics are small, our result
suggests an increase of metallicities from $z>3$ to present
times. Previous compilations of Zn measurements at all redshifts have
shown that the column density weighted metallicity does not evolve
with time (Pettini \etal\ 1997, Vladilo \etal\ 2000, Prochaska \&
Wolfe 2002). These authors argue however that this cannot be directly
interpreted as a lack of evolution of the cosmic metallicity of DLAs
since measurements are still affected by low number statistics and
possible bias against high column density, high metallicity absorbers
(as first pointed out by Fall \& Pei 1993 and Boiss\'e \etal\
1998). The presence of this bias is controversial at the moment
(Ellison \etal\ 2001, Prochaska \& Wolfe 2002, Petitjean \etal\ 2002)
and many more measurements of undepleted metals in quasar absorbers
are needed at $z>3$ before any conclusion can be drawn.

\section{Acknowledgments}
We are grateful to C\'edric Ledoux for reducing the spectrum of BR
1117$-$1329 presented here and to an anonymous referee for providing
very constructive comments. C. P. thanks Bob Carswell and Miriam
Centuri\'on for help with VPFIT and FITLYMAN respectively and Giovanni
Vladilo and Paolo Molaro for comments on an earlier version of this
manuscript. This work was supported in part by the European
Communities RTN network "The Physics of the Intergalactic
Medium". P.P.J. and R. S. gratefully acknowledge support from the
Indo-French Centre for the Promotion of Advanced Research (Centre
Franco-Indien pour la Promotion de la Recherche Avanc\'ee) under
contract No. 1710-1.

% The phrase \cite{Bai92} produces (Bailyn 1992).
% In the phrase \citeasnoun{Bai95} Bailyn et al. (1995) appear as a noun.
% Affixes (e.g. Barnes et al. 1976) are produced by the phrase
% \citeaffixed{Barnes et al. 1976}{e.g.}.
% Other options of the harvard package, e.g. \citeyear, are not
% reproduced in New Astronomy.

\end{document}